\documentclass[11pt]{article} \textheight=22 true cm
\usepackage{subfigure}
\usepackage{graphicx}
\usepackage{amsmath}
\usepackage{amssymb}
\usepackage{color}
\DeclareMathOperator{\csch}{csch}

\begin{document}
\begin{center}
{\Large\bf Studies of the Inhomogeneous Cosmology in Higher Dimensional space-time with a Cosmological Constant}\\[15 mm]
D. Panigrahi\footnote{ Netaji Nagar Day College, 170/436 N.S.C. Bose Road, Regent Estate, Kolkata  700092, INDIA
\emph{and also} Relativity and Cosmology Research Centre, Jadavpur
University, Kolkata - 70032, India , e-mail:
dibyendupanigrahi@yahoo.co.in, dpanigrahi@nndc.ac.in }
  and S. Chatterjee\footnote{ New Alipore College (Retd.), Kolkata - 700053, India \emph{and
also} Relativity and Cosmology Research Centre, Jadavpur
University,
Kolkata - 700032, India, e-mail : chat\_sujit1@yahoo.com } \\[10mm]

\end{center}

\begin{abstract}
We have studied the inhomogeneous cosmology in Kaluza-Klein spacetime with a positive cosmological constant in a dust dominated era ($p = 0$). Depending on the integration constant we have derived two types of solutions. The dimensional reduction  of extra dimensional scale factor is possible due to inhomogeneity depending on the curvature of the metric for positive cosmological constant for all solutions. The high value of entropy in present observable universe  and the possible matter leakage in $4D$ world due to reduction of extra dimension are also discussed. Our solutions show the early deceleration and late accelerating nature of the universe. Findings are verified by the wellknown Raychaudhuri equation.

\end{abstract}

\bigskip
\section{ Introduction}

It is wellknown from Kaluza-Klein (KK) theories that electromagnetism and gravity can be unified from Einstein's field equations if we add an extradimension to the theory of gravitation. Chodos and Detweiler~\cite{cd} have first shown  that the four dimensional  vacuum solutions can be extended to five dimensions where the metric may be chosen in such a way that the dimensional reduction takes place in the extra space, whereas the usual three space expands indefinitely. In a pioneering work Tosa~\cite{tosa} studied the higher dimensional cosmology in homogeneous spacetime with a cosmological constant. He observed that the dimensional reduction takes place for extra dimension in an anti de-Sitter spacetime while usual $3D$ space expands indefinitely with time. Later Banerjee \textit{et al}~\cite{ban1, ban2} have extended the idea to the inhomogeneous spacetime also and shown that the extra space contracts even with a positive cosmological constant. In this context one may recall that homogeneous spacetime have been extensively discussed but little attention has been paid so far to inhomogeneous spacetime in the field.

However, the observational findings of both Cosmic Background Explorer (COBE)~\cite{sm} and later Planck Satellite data ~\cite{oo} favour the inhomogeneity in our early universe. While homogeneous higher dimensional model has been extensively studied, its inhomogeneous generalisation has been paid scant attention in the literature so far. To cite a few we refer to the recent works of Bronnikov et al~\cite{bro}, Bojowald et al ~\cite{bo}, Brassel et al ~\cite{bra} etc in this field.   On the other hand if higher dimensional spacetime is to be seriously taken as an alternative approach to some real physics, it is only in the context  of the early phase of cosmological evolution that inclusion of an extra space assumes a particular significance. However, our observable universe is manifestly  isotropic, homogeneous and flat, although any satisfactory explanation of these observed properties continues to be evasive.

On the other hand if we follow the Einstein's theory of General Relativity and FRW type of model then as standardised candles type Ia supernovae pointed out to ~\cite{res} a universe  which is undergoing accelerated expansion with about $5\%$ baryonic matter of the total energy. Later data from CMBR probes~\cite{spe} also lead to the same finding. This has  led  a good majority of cosmology community to investigate the apparent  cause of the accelerated expansion. The puzzling  question in this field is the possible identification of the processes likely to be responsible for starting the late acceleration.
In this context it may be pointed out that there are several approaches like $f(R)$ gravity ~\cite{st}, Chaplygin type of gas as matter field~\cite{ka, dp0}, $\Lambda$CDM model~\cite{dav}, Higher dimensional spacetime~\cite{dp1} etc. to explain the above phenomena. However,  the $\Lambda$CDM model~\cite{dav} is so far the best fit to explain observational findings.   It is generally  accepted that the dark energy~\cite{mae} candidate $\Lambda$ in Einstein’s field equations is responsible for the cosmic acceleration of the universe.  On the other hand, there is a huge difference between the observed value and the value obtainable from particle physics.

In this paper we have considered an inhomogeneous five-dimensional model with a positive cosmological constant for dust dominated universe ($p = 0)$ where eventually the extra dimensional pressure ($p_5$) automatically vanishes in our solution. So there is no apparent excitation in the extra dimension - this position is assumed in Lee and Lee's work ~\cite{lee} also. Here the three dimensional metric is spherically symmetric and is  considered to be a function of time only whereas the extra space is the function of both time and radial coordinate. This work is the generalisation of Banerjee \textit{et al}~\cite{ban1,ban2}. There some important differences are observed depending upon the relation between the integration constants. The main difference is that the exponential solutions of three spaces as well as the extra space are observed in the work of  Banerjee \textit{et al}~\cite{ban1,ban2}. But in this work, adjusting the expressions of the integration constants, we have got another two hyperbolic solutions of three spaces and extra space for positive cosmological constant. Three space expands indefinitely with time and shows acceleration at the late stage of the universe. We have calculated \textit{flip} time also. But the extra space starts from infinity at $t = 0$ and then it shows dimensional reduction with one minima and after that it expands indefinitely depending upon the curvature of the spacetime. We have also discussed the entropy generation of three space where it is shown that the high value of entropy per baryon observable at present universe due to dimensional reduction of extra space which is also studied earlier by Alvarez and Gavela~\cite{al}. It is also observed that as the extra dimension shrinks, the total $5D$ world matter is conserved, and there will be matter leakage from the internal space onto \textit{effective} $4D$ world for $r$-constant hypersurface. Finally we examined the expressions of matter density  obtained from different models with those obtained from the Raychaudhuri equation~\cite{ray1,ban3,el} in 5-dimensional spacetime with positive cosmological constant and next we have determined the deceleration parameter from the Raychaudhuri equation.

In section 2 we derive the solutions using field equations with the help of the equation of state ($p = 0$). Depending on the value of the integration constant $k$ in equation \eqref{eq:14a}, we get two types of solutions denoted by Model-1 ($k = 1$) \& Model-2 ($k = 0$). Again in Model-1 we have derived  two  hyperbolic solutions of $e^{\mu}$ and in Model-2  we get exponential solution which was discussed in the previous work of Banerjee \textit{et al}~\cite{ban1}. In section 3, the analysis of  the findings  using the solutions found in section 2 are investigated in some detail. The entropy generation and matter leakage were also briefly  studied in section 4. Next accelerating nature of our solution is investigated. For completeness, we confirm  our inferences obtained from different models with those obtained from the Raychaudhuri equation ~\cite{ray1,ban3,el}, and this paper ends with a discussion in section 5.

\section{ Field Equations and their solutions}

We consider a five dimensional spacetime line element is as
\begin{equation}\label{eq:1}
  ds^{2}= dt^{2}- e^{2\lambda}\left(dr^2 + r^2 d\theta^2 + r^2 sin^2 \theta d \phi^2 \right) -e^{2\mu} dy^2
\end{equation}
where $\lambda = \lambda(t)$ and $\mu = \mu(r,t)$.

Here we assume that the physical three-space is flat and homogeneous while inhomogeneity is present in the extra space only. The stress-tensor, $T_{ij}$ \textbf{(here $i, j = 0, 1, 2, 3, 4$)} in comoving coordinate will  be

\begin{equation}\label{eq:2}
T^0_0 = \rho + \Lambda, T^1_1 = T^2_2 = T^3_3 = - p + \Lambda, T^4_4 = - p_5 + \Lambda
\end{equation}

where $ \Lambda $ denotes the cosmological constant. Here the pressure corresponding to the fifth dimension is denoted as $p_5$.

The field equations are given below

\begin{equation}\label{eq:3}
G_{01} =  \dot{\mu'} + \dot{\mu} \mu' - \dot{\lambda} \mu' = 0
\end{equation}

\begin{equation}\label{eq:4}
G^1_1 = e^{-2\lambda} \frac{2\mu'}{r} -\left(2\ddot{\lambda} + 3\dot{\lambda}^2 + \ddot{\mu} + \dot{\mu}^2 + 2\dot{\lambda}\dot{\mu} \right) = p - \Lambda
\end{equation}

\begin{equation}\label{eq:5}
G^2_2 = G^3_3 = e^{-2\lambda}\left(\mu'' + \mu'^2 + \frac{\mu'}{r}\right) -\left(2\ddot{\lambda} + 3\dot{\lambda}^2 + \ddot{\mu} + \dot{\mu}^2 + \dot{\lambda}\dot{\mu} \right) = p - \Lambda
\end{equation}

\begin{equation}\label{eq:6}
G^4_4 = - 3\left(\ddot{\lambda} + 2\dot{\lambda^2} \right) = p_5 - \Lambda
\end{equation}

\begin{equation}\label{eq:7}
G^0_0 = e^{-2\lambda}\left(\mu'' + \mu'^2 + 2\frac{\mu'}{r}\right) -3\left(\dot{\lambda}^2 +  \dot{\lambda}\dot{\mu} \right) = - \rho - \Lambda
\end{equation}

where a dot and prime overhead denote differentiation with respect to time and radial coordinate respectively.
Again, from the time component of the Bianchi identity we get

\begin{equation}\label{eq:8}
\dot{\rho} + 3\left(p + \rho \right)\dot{\lambda} + \left(p_5 + \rho \right)\dot{\mu} = 0
\end{equation}

We assume an equation of state of the form $p = (\gamma - 1) \rho \neq p_5$. We further assume that $\gamma =1 $ for the dust case only.

On integration equation \eqref{eq:3} gives

\begin{equation}\label{eq:9}
e^{\mu} = A(r) e^{\lambda} + \alpha(t)
\end{equation}

where $A$ and $\alpha$ are arbitrary functions of space and time respectively.
Again from equations \eqref{eq:4} and \eqref{eq:5}, we get

\begin{equation}\label{eq:10}
e^{\mu} = -a(t) br^2 + \alpha(t)
\end{equation}

where $a$ is again an arbitrary function of time alone and $b$ is an arbitrary constant of integration.
Now comparing equations \eqref{eq:9} and \eqref{eq:10} we get

\begin{equation}\label{eq:11}
e^{\mu} = - br^2 e^{\lambda} + \alpha(t)
\end{equation}

Using here the equations \eqref{eq:4} and \eqref{eq:11}, we get the following pair of equations:

\begin{eqnarray}
\ddot{a}a + \dot{a}^2 - \frac{\Lambda a^2}{3} &=& 0   \label{eq:12}\\
a \ddot{\alpha} + 2\dot{a}\dot{\alpha} + \left(2 \ddot{a} + \frac{\dot{a}^2}{a}\right)\alpha - \Lambda a \alpha &=& -4b  \label{eq:13}
\end{eqnarray}

Now integrating \eqref{eq:12} we obtain

\begin{equation}\label{eq:14}
a(t) = \left(\frac{3}{2 \Lambda} \right)^{\frac{1}{4}} \left( \frac{e^{4\sqrt{\frac{\Lambda}{6}}t}- k}{e^{2\sqrt{\frac{\Lambda}{6}}t}}\right)^{\frac{1}{2}} = \frac{1}{\sqrt{2 \beta}} \sqrt{e^{-2\beta t} \left(e^{4 \beta t} - k \right)}, ~~~~~~~~ \Lambda>0
\end{equation}
where $\beta^2 = \frac{\Lambda}{6}$, $k$ is the integration constant and obviously $k \leq 1$.

Using equation \eqref{eq:13} and equation \eqref{eq:14} we get

\begin{equation}\label{eq:14a}
\alpha(t) = \frac{\sqrt{2}b e^{2 \beta t}\sqrt{e^{-2\beta t} \left(e^{4 \beta t} - k \right)}} {\beta^{\frac{3}{2}} \left(e^{4\beta t }-k\right)} + \frac{c_1 e^{3\beta t}}{4 \beta\sqrt{ e^{4 \beta t} - k }}+
\frac{c_2 e^{-\beta t}}{\sqrt{ e^{4 \beta t} - k }}
\end{equation}

To get simplified solutions we consider $k = 1$  \& $k = 0$. Now we calculate different types of solutions as given below.

\subsection{ Model 1 ($k = 1$)}

 Now substituting $k = 1$ in the above solutions.
From equation \eqref{eq:14} we obtain,

\begin{equation}  \label{eq:15}
a(t) = \frac{1}{\sqrt{ \beta}} \sinh^{\frac{1}{2}}  2\beta t
\end{equation}

From \eqref{eq:14a} we get the following equation

\begin{equation}  \label{eq:17}
\alpha(t) = \left( \frac{c_1 }{4 \beta} e^{2 \beta t} + c_2 e^{- 2 \beta t} \right) \frac{\sinh^{-\frac{1}{2}} 2 \beta t}{\sqrt{2}} + \frac{b }{\beta^{\frac{3}{2}}}\sinh^{-\frac{1}{2}} 2 \beta t
\end{equation}

Again to  get more simplified solutions we consider two relations between the constants:

\subsubsection{ Case I}

Let $\frac{c_1}{4 \beta} = - c_2 = \frac{c}{\sqrt{2}}$,

using this relation in equation \eqref{eq:14a}, we obtain

\begin{equation}  \label{eq:18}
\alpha(t) = c \sinh^{\frac{1}{2}} 2 \beta t + \frac{b }{\beta^{\frac{3}{2}}}\sinh^{-\frac{1}{2}} 2 \beta t
\end{equation}

and $e^{\mu}$ will be

\begin{equation}  \label{eq:19}
e^{\mu} = \left(c - \frac{br^2}{\sqrt{\beta}} \right) \sinh^{\frac{1}{2}} 2 \beta t + \frac{b}{\beta^{\frac{3}{2}}}\sinh^{-\frac{1}{2}} 2 \beta t
\end{equation}

\subsubsection{ Case II}

Let $\frac{c_1}{4 \beta} = c_2 = \frac{c}{\sqrt{2}}$,

and the equation \eqref{eq:14a} reduces to

\begin{equation}  \label{eq:20}
\alpha(t) = c \cosh 2\beta t \sinh^{-\frac{1}{2}} 2 \beta t + \frac{b }{\beta^{\frac{3}{2}}}\sinh^{-\frac{1}{2}} 2 \beta t
\end{equation}

In this case the expression of  $e^{\mu}$ will be
\begin{equation}  \label{eq:21}
e^{\mu} = \left(c \coth 2 \beta t - \frac{br^2}{\sqrt{ \beta}} \right) \sinh^{\frac{1}{2}} 2 \beta t + \frac{b }{\beta^{\frac{3}{2}}}\sinh^{-\frac{1}{2}} 2 \beta t
\end{equation}

For both the cases

\begin{equation}  \label{eq:22}
e^{\lambda} = \frac{1}{\sqrt{ \beta}} \sinh^{\frac{1}{2}}  2\beta t
\end{equation}

\subsection{ Model 2 ($k = 0$)}

From equation \eqref{eq:14} we obtain

\begin{equation}\label{eq:23}
a(t) = \left(\frac{3}{2\Lambda} \right)^{\frac{1}{4}} e^{\sqrt{\frac{\Lambda}{6}}t} = \frac{1}{\sqrt{2\beta}}e^{\beta t}
\end{equation}

Using equation \eqref{eq:14a}  we obtain the solution

\begin{equation}\label{eq:24}
\alpha(t) = c e^{\beta t} - Be^{-3\beta t} + \frac{b \sqrt{2} e^{-\beta t}}{ \beta^{\frac{3}{2}}}
\end{equation}

So the metric \eqref{eq:1} is now expressed in the form:

\begin{eqnarray}
e^{\lambda} &=& \frac{1}{\sqrt{2\beta}}e^{\beta t}  \label{eq:25}  \\
e^{\mu} &=& \left(c -   \frac{br^2}{\sqrt{2 \beta}}  \right)e^{\beta t} - Be^{-3\beta t} + \frac{b \sqrt{2} e^{-\beta t}}{ \beta^{\frac{3}{2}}}              \label{eq:26}
\end{eqnarray}

\section{Dynamical behaviour }

First we have to check if our spacetime contains any geometric singularity apart from the wellknown bigbang one at $t =0$. We calculate the well known Kreschmann scalar for the metric \eqref{eq:1}    given by

\begin{equation}  \label{eq:27}
R^{ijkl}R_{ijkl} = 3 \lambda^4 + 3 \left( \ddot{\lambda} + \dot{\lambda}^2 \right)^2 + \left(\ddot{\mu} + \dot{\mu}^2 \right)^2 + 2 \left(\dot{\mu} \dot{\lambda} - e^{- 2\lambda} \frac{\mu'}{r} \right)^2
\end{equation}

as $t \rightarrow 0$, the invariant diverges but it is regular at $r = 0$, since $\frac{\mu'}{r}$ is regular there. So there is no spatial singularity in our inhomogeneous model which may be identified with the centre of the fluid distribution as in many other inhomogeneous distribution.

Now the 4-space curvature of the t-constant hyper-surface  $R^{*(4)}$ for the line element \eqref{eq:1}, from the expression ~\cite{roy} is

\begin{equation}  \label{eq:28}
R^i_i = R^{*(4)} + \dot{\theta} + \theta^2 - 2 \omega^2 + u^i_{;i}
\end{equation}
Here $R^i_i$ is known as Ricci scalar; $i = 1,2,3,4$ for $5$-dimensional cases and  $\theta$ is the expansion scalar and the last two terms refer to the vorticity and acceleration. After a long but straight forward calculation we get

\begin{equation}  \label{eq:29}
R^{*(4)} = - e^{-2\lambda} \frac{6 \mu'}{r}
\end{equation}

\subsection{ Model 1 }

Here we have two solutions of $e^{\mu}$ and hence our analysis will be of two types. So first we discuss the results of solution 1.

\subsubsection{ Case I}

Using the results \eqref{eq:7},\eqref{eq:19} and \eqref{eq:22} the matter density will be

\begin{equation}  \label{eq:30}
\rho = \frac{6 \beta^{2}\left(c - \frac{br^2}{\sqrt{ \beta}}\right) \csch 2 \beta t}{\left(c - \frac{br^2}{\sqrt{\beta}}\right) \sinh 2 \beta t + \frac{b}{\beta^{\frac{3}{2}}}}
\end{equation}

\begin{figure}[ht]
\centering \subfigure[ $e^{\lambda}$ vs $t$ using equation \eqref{eq:22} ]{
\includegraphics[width= 6 cm]{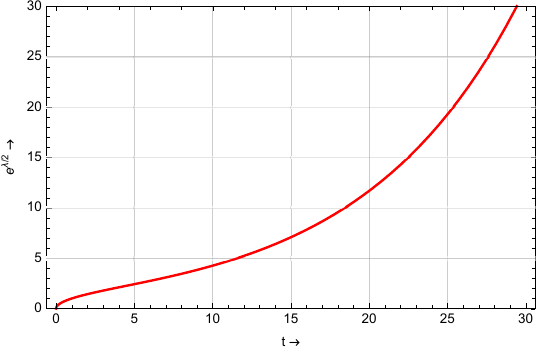}\label{fig:rt}
\label{fig:subfig1} } ~~~\subfigure[   $\rho$ vs $t$ using equation \eqref{eq:30}
 ]{
\includegraphics[width= 6 cm]{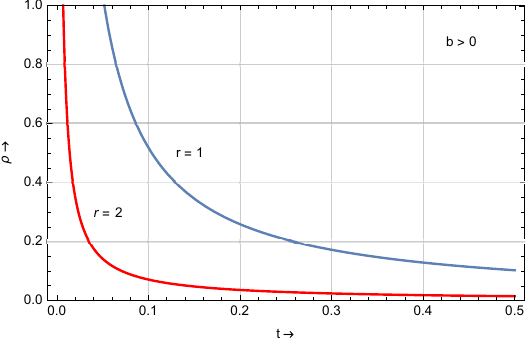}\label{fig:rhot1} }\label{fig:fig1}~~~~~~~~~~~\caption[Optional
caption for list of figures]{\emph{  $\left(c - \frac{br^2}{\sqrt{ \beta}}\right) > 0$   }}
\end{figure}

We have studied here the case of $\Lambda > 0$. To get the positive matter density $\left(c - \frac{br^2}{\sqrt{ \beta}}\right)$ must be positive. Initially ( i.e., $t = 0$), the matter density was infinite as expected and it reduces to zero with time which is shown in the fig. \ref{fig:rhot1}.  Here we find that the  extra dimensional pressure  $p_5 = 0$ automatically. Further for $b > 0$, the mass density vanishes at the radial coordinate $\sqrt{\frac{c}{b}}\left( \beta\right)^{\frac{1}{4}}$and this may be taken as a natural choice for our coordinate boundary.

Using equations \eqref{eq:19}, \eqref{eq:22} and  \eqref{eq:29}, the 4-space curvature for this solution is given by

\begin{equation}  \label{eq:31}
R^{*(4)}  = \frac{12 \sqrt{\beta }b}{\left(c - \frac{br^2}{\sqrt{ \beta}} \right) \sinh^{\frac{1}{2}} 2 \beta t + \frac{b }{\beta^{\frac{3}{2}}}}
\end{equation}

So the arbitrary constant $b$ determines both the measure and nature of curvature of the 4D space. Depending on the signature of $b$, we get positive, negative or flat space according as $b >, < , 0$.  Here it is seen that $R^{*(4)} \rightarrow 0$ as $ t \rightarrow \infty$ implying a  flat surface.

The shear scalar for our model is

\begin{equation}  \label{eq:32}
\sigma^2  = \frac{3}{8}\left(\dot{\mu} - \dot{\lambda} \right)^2 = \frac{3}{2\beta} b^2 \coth^2 2 \beta t
 \csch 2 \beta t ~ e^{- 2 \mu}
\end{equation}

\begin{figure}[ht]
\begin{center}
  \includegraphics[width=8 cm]{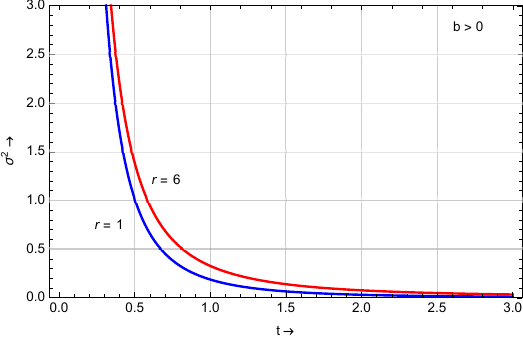}
  \caption{
  \small\emph{The variation of $\sigma^2$ with $t$ using equation \eqref{eq:32}}\label{fig:st1}
    }
\end{center}
\end{figure}

AS $t \rightarrow \infty$, $\sigma^2 $ vanishes as expected. Here radial coordinate also affects the value of shear (see fig. \ref{fig:st1}).

The total 5D volume is given by

\begin{equation}  \label{eq:33}
V = e^{3 \lambda} e^{\mu} = \left(\frac{1}{ \beta} \right)^{\frac{3}{2}}
\sinh 2 \beta t \left\{\left(c - \frac{b r^2}{\sqrt{ \beta}} \right) \sinh 2 \beta t  + \frac{b }{\beta^{\frac{3}{2}}} \right\}
\end{equation}

\begin{figure}[ht]
\centering \subfigure[ $V$ vs $t$ using equation \eqref{eq:33} ]{
\includegraphics[width= 6 cm]{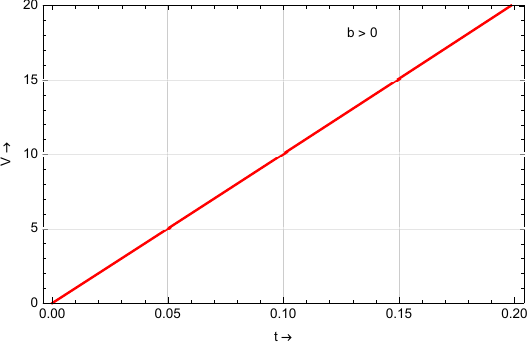}
\label{fig:Vt1} } ~~~\subfigure[   $e^{\mu}$ vs $t$ using equation \eqref{eq:19}
 ]{
\includegraphics[width= 6 cm]{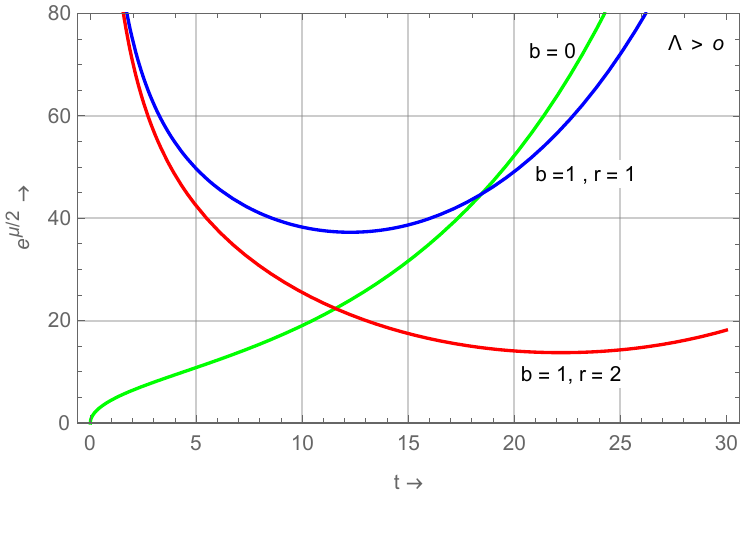}

 \label{fig:At1} }\label{fig:fig3}~~~~~~~~~~~\caption[Optional
caption for list of figures]{\emph{ The condition is $\left(c - \frac{br^2}{\sqrt{ \beta}}\right)$ and $b > 0$  }}
\end{figure}

The equation \eqref{eq:33} shows that 5D volume expands monotonically for $b > 0$  as shown in the fig. \ref{fig:Vt1} though extra dimension has one minima at

\begin{equation}  \label{eq:34}
t_{min} = \frac{1}{2 \beta} \sinh^{-1} \left\{ \frac{b}{\beta^{\frac{3}{2}} \left( c - \frac{b r^2}{\sqrt{ \beta}}\right)}\right\}
\end{equation}

It is found that dimensional reduction occurs due to inhomogeneity. Further it is noticed that the constant $ b$ is the measure of curvature and it accounts for inhomogeneity also. So it may be pointed out that the curvature is the  cause of dimensional reduction of extra space for $\Lambda > 0$. It is also seen that when $b = 0$, the inhomogeneity of extra space disappears and it expands indefinitely which is shown in the fig \ref{fig:At1}. In this context, we refer to  the work of Tosa ~\cite{tosa} where dimensional reduction is not possible for homogeneous model when $\Lambda > 0$. So $b$ is the factor which changes the total scenario. As the inhomogeneity parameter $b$ tends to zero the coordinate boundary shifts more and more from the point of symmetry and the cosmology reduces to a $5D$ homogeneous model. It is to be noted that the dimensional reduction is affected by the spatial co-ordinate $r$ also. Here 3D scale factor expands indefinitely.

\subsubsection{ Case II}

Now using \eqref{eq:21}, \eqref{eq:22} and \eqref{eq:29} we get the 4-space curvature for this solution as

\begin{equation}  \label{eq:35}
R^{*(4)}  = \frac{12 \sqrt{ \beta} b}{\left(c \coth 2 \beta t - \frac{br^2}{\sqrt{ \beta}} \right) \sinh 2 \beta t + \frac{b }{\beta^{\frac{3}{2}}}}
\end{equation}

which shows that $b$ is the measure of curvature. The shear scalar for this case is given by

\begin{equation}  \label{eq:36}
\sigma^2 = \frac{3}{2} \beta^2 \left(c + \frac{b }{\beta^{\frac{3}{2}}} \cosh 2 \beta t \right)^2 \csch^3 2 \beta t~   e^{-2\mu}
\end{equation}

\begin{figure}[ht]
\centering \subfigure[ $\sigma^2$ vs $t$ using equation \eqref{eq:36} ]{
\includegraphics[width= 6 cm]{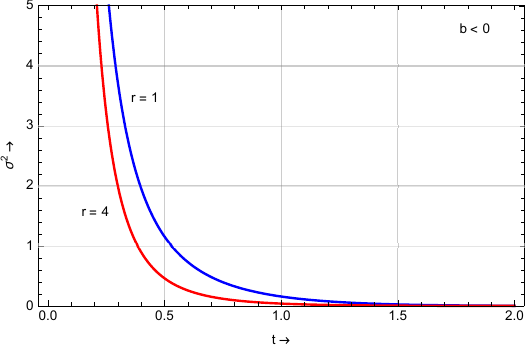}
\label{fig:st2} } ~~~\subfigure[   $\rho$ vs $t$ using equation \eqref{eq:37}
 ]{
\includegraphics[width= 6 cm]{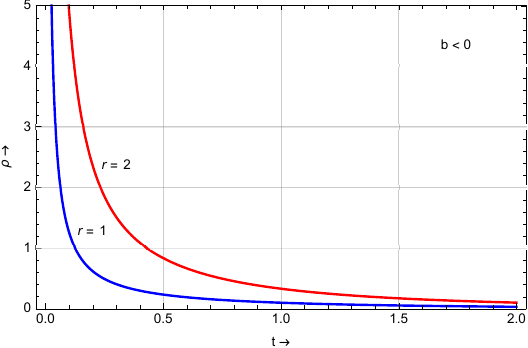}\label{fig:rhot2} }\label{fig:fig4}~~~~~~~~~~~\caption[Optional
caption for list of figures]{\emph{ $  c > \frac{b }{\beta^{\frac{3}{2}}} $  }}
\end{figure}

AS $t \rightarrow \infty$, $\sigma^2 $ vanishes as expected. Here radial coordinate also affects the value of shear (see fig. \ref{fig:st2}).

\begin{figure}[ht]
\centering \subfigure[ $V$ vs $t$ using equation \eqref{eq:38} ]{
\includegraphics[width= 6 cm]{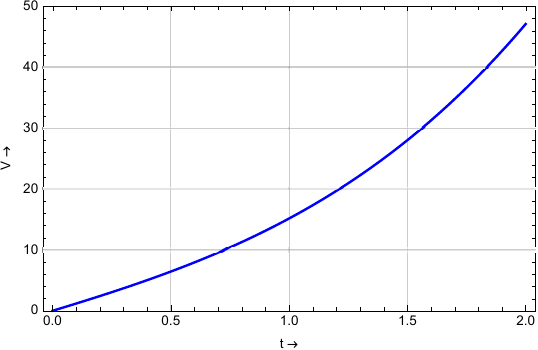}
\label{fig:Vt2} } ~~~\subfigure[   $e^{\mu}$ vs $t$ using equation \eqref{eq:39}
 ]{
\includegraphics[width= 6 cm]{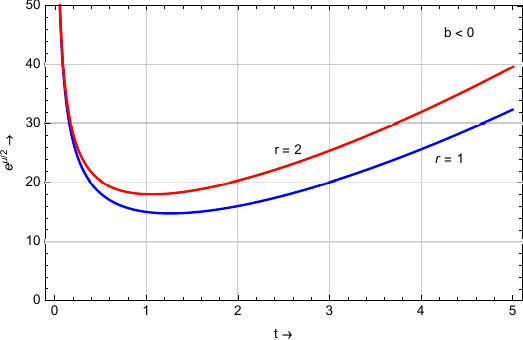}

 \label{fig:At2} }\label{fig:fig5}~~~~~~~~~~~\caption[Optional
caption for list of figures]{\emph{ The condition is $c > \frac{d }{\beta^{\frac{3}{2}}}$ with negative curvature  }}
\end{figure}

To get the positive matter density we have to assume that the constant $b$ should be negative for this case. Let $d = - b$ and in what follows we are dealing with a negative curvature model. Now using the equations \eqref{eq:7}, \eqref{eq:21} and \eqref{eq:22} we obtain the matter density as

\begin{equation}  \label{eq:37}
\rho =  \frac{6\beta^{\frac{3}{2}} d r^2 \csch 2 \beta t}{\left(c \cosh 2 \beta t + \frac{dr^2}{\sqrt{\beta}} \sinh 2 \beta t \right) - \frac{d }{\beta^{\frac{3}{2}}}}
\end{equation}

which forces  $c > \frac{d}{\beta^{\frac{3}{2}}}$ for physically realistic mass density. In this case the matter density decreases to zero from infinity as universe expands for negative curvature.  Here the fifth dimensional pressure $p_5 = 0$. Now the 5D volume will be

\begin{equation}  \label{eq:38}
V  = \frac{1}{ \beta^{\frac{3}{2}}}\left\{\left(c \coth 2 \beta t + \frac{d r^2}{\sqrt{ \beta}} \right) \sinh^2 2 \beta t  - \frac{d }{\beta^{\frac{3}{2}}}\sinh 2 \beta t \right\}
\end{equation}

which expands indefinitely as shown in the fig. \ref{fig:Vt2}.

Now the extra dimension becomes

\begin{equation}  \label{eq:39}
e^{\mu} = \left(c \coth 2 \beta t + \frac{dr^2}{\sqrt{ \beta}} \right) \sinh^{\frac{1}{2}} 2 \beta t - \frac{d }{\beta^{\frac{3}{2}}}\sinh^{-\frac{1}{2}} 2 \beta t
\end{equation}

The desirable feature of dimensional reduction is possible here but it has one minima and after that extra dimension expands indefinitely for negative curvature with $\Lambda > 0$ (see fig. \ref{fig:At2}), when we are considering the inhomogeneity of extra space only. The dimensional reduction is  affected by the spatial coordinate $r$ also.

\subsection{Model 2}

Using equations \eqref{eq:7}, \eqref{eq:25} and \eqref{eq:26}we get the expression of matter density as

\begin{equation}  \label{eq:40}
\rho = \frac{12 B \beta^2}{\left(c -   \frac{br^2}{\sqrt{2 \beta}}  \right)e^{4\beta t} - B + \frac{b \sqrt{2} e^{2\beta t}}{ \beta^{\frac{3}{2}}}}
\end{equation}

where $\left(c -   \frac{br^2}{\sqrt{2 \beta}}  \right) >0$. We have discussed here dust case only, i.e., $p =0$. But the fifth dimensional pressure $p_5 $ vanishes automatically.

\begin{figure}[ht]
\centering \subfigure[ $e^{\mu}$ vs $t$ using equation \eqref{eq:26} ]{
\includegraphics[width= 6 cm]{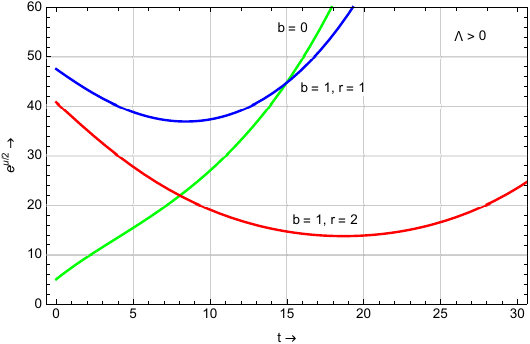}
\label{fig:At3} } ~~~\subfigure[   $\rho$ vs $t$ using equation \eqref{eq:40}
 ]{
\includegraphics[width= 6 cm]{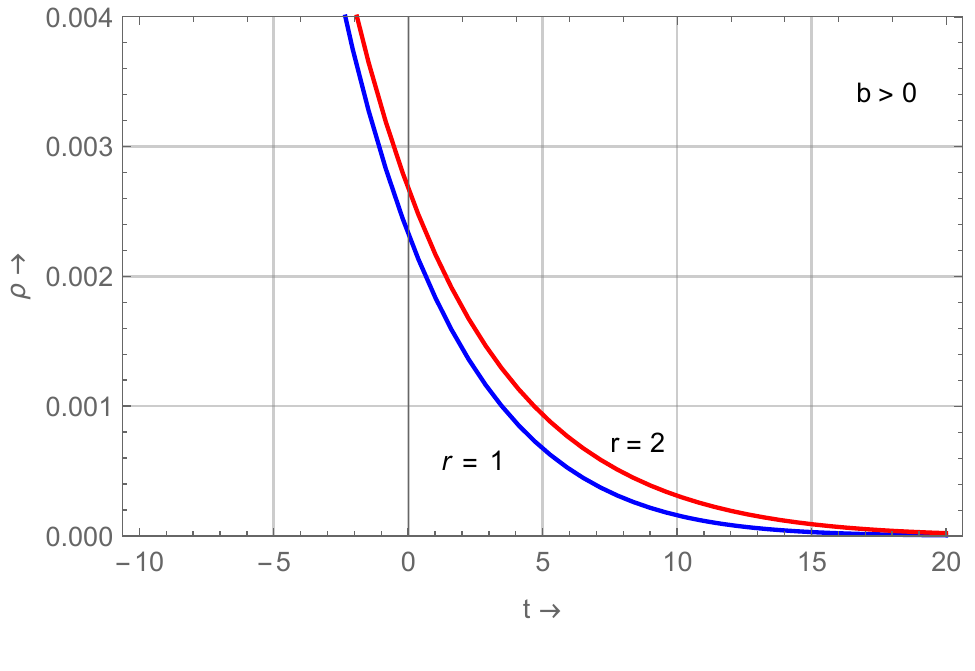}\label{fig:rho0} }\label{fig:fig6}~~~~~~~~~~~\caption[Optional
caption for list of figures]{\emph{ $\left(c - \frac{br^2}{\sqrt{2 \beta}}\right)> 0$  }}
\end{figure}

Using equations \eqref{eq:25}, \eqref{eq:26} and \eqref{eq:29}, the 4-space curvature for this solution is given by

\begin{equation}  \label{eq:41}
R^{*(4)} = \frac{12 \sqrt{2 \beta} b}{  \left( c - \frac{br^2}{\sqrt{2\beta}}\right)e^{2 \beta t} -  Be^{-2\beta t} + \frac{b \sqrt{2} }{ \beta ^{\frac{3}{2}}}}
\end{equation}

Here arbitrary constant $b$ also determines the measure of curvature of the 4D space(t-costant hypersurface). In this model, if $b > 0$, \textit{i.e.}, for positive curvature, there is no singularity at any finite time in the evolutionary history of universe. $\rho$ is finite and positive at $t \rightarrow  - \infty$ (see fig. \ref{fig:rho0}). Here the three space  expands indefinitely with time. The volume also increases from a finite magnitude without any bounce at any stage in future. The dimensional reduction is possible in this case, however, there will  only be a one minima in $e^{\mu}$ and after that it expands indefinitely as shown in the fig. \ref{fig:At3}. To avoid this shortcoming, \textit{i.e.}, the expansion of extra dimension, we assume that some stabilising mechanism (which is beyond the scope in this article) stabilises it at a certain minimum value and thereafter no expansion is possible.

 It is interesting to note that when $b <0$ the extra dimension contracts indefinitely and we get the desirable feature of dimensional reduction. Moreover, since we are considering an inhomogeneous model, the dimensional reduction takes place at different $r$ at different epochs.
This was studied in our previous work ~\cite{ban1,ban2} in detail.
 It may not be out of place here to refer to the work of Beloborodov et al~\cite{bel} where it is concluded that for a homogeneous model exponential expansion of the usual scale is not compatible with dimensional reduction of the extra scale. It follows from our work that the above conclusion is restricted to homogeneous model only for positive curvature and positive cosmological constant. When the inhomogeneity parameter $b$ as well as curvature associated with our model vanishes, our spacetime becomes homogeneous  and extra space show indefinite expansion which is in line with the findings of Beloborodov \textit{et al} ~\cite{bel}. So it is the inhomogeneity parameter that makes the all differences for positive curvature with $\Lambda > 0$.

 Relevant to mention  an earlier work of Bronnikov and Rubin ~\cite{bro1} in the Starobinsky frame work where it is shown that $f(R)$ gravity may account for spontaneous compactification simultaneously with expansion of $3D $ space. Interestingly we also get similar result in our Inhomogeneous Higher Dimensional model although several workers in the field ~\cite{cd,sah} got similar results in the past in homogeneous model.

 \section{Some physical consequences }

 In this section we have investigated the nature of our solutions and compared the findings from these to wellknown Raychaudhuri equation. Two view points exist about the role of extra dimensions in physics. While a section of workers look at as a mathematical artifact to reformulate physical laws in a new form, the other section take it seriously to address physical problems confronting us and try to solve. Taking the second view point we try to focus our attention to two vexed problems of current cosmology namely big bang singularity and also the entropy per baryon currently observable.

 \subsection{ Entropy generation }

 We know that high value of entropy per baryon is observed in the present universe.  Now our assumptions of the laws of thermodynamics in 4D approach may be generalised to higher dimensional spacetime also  and that interactions at the initial stage of the expansion of our cosmos led to the thermodynamic equilibrium, the total entropy conserves. The energy density is expressed as $\rho = \frac{U}{V}$, where $U$ and $V$ are the internal energy and volume filled by the fluid respectively. Since we are dealing with a dust dominated universe (i.e., $p = p_5 = 0$),  we may write

\begin{equation}  \label{eq:42}
dU = TdS = V d \rho + \rho dV
\end{equation}

where $S$ is the entropy at temperature $T$.

If we define the effective entropy of three space is $S_3$, then from equation \eqref{eq:42} we get

\begin{equation}  \label{eq:43}
T dS_3 = e^{3\lambda} d\rho + \rho d(e^{3\lambda})
\end{equation}

where, $e^{3\lambda}$ is the three space volume. Now the equations \eqref{eq:8} and \eqref{eq:43} yield

\begin{equation}  \label{eq:44}
T \dot{S_3} = -\rho e^{3\lambda}\dot{\mu}
\end{equation}

The above equation is very significant in the sense that the \textit{effective} entropy in three space increases as a result of the decrease of the extra dimensional space. Since $\rho > 0$ and $\dot{S_3} > 0$ when $\dot{\mu} < 0$. So we may conclude that the very high value of entropy in our observable universe might be due to the compactification of the  extra-dimensional space.

\subsection{ Matter leakage}

From another view point, one may look at the conservation relation \eqref{eq:8}. To avoid big bang singularity, Bondi and Gold ~\cite{bon} suggested a steady state model with the help of continuous creation of matter without offering any dynamical mechanism. Later Hoyle and Narlikar ~\cite{nar} introduced an external “creation field” in their C-field theory for matter creation violating conservation principle. There is also another view point in the case of $5D$ world where the conservation principle is strictly valid. For our dust case in a $r$-constant hypersurface, equation \eqref{eq:8}  reduces to

\begin{equation}  \label{eq:45}
\rho e^{3\lambda}e^{\mu} = constant = M_5
\end{equation}

where $M_5$ may be identified with the total mass in the $5D $ world which strictly conserves for $r$-constant hypersurface.
Now we study the $M_5$ based on our findings from different models discussed previously as given below,

(i) Using equations \eqref{eq:19}, \eqref{eq:22} and \eqref{eq:30} we finally get

\begin{equation}  \label{eq:45a}
M_5 = \frac{3}{2} \sqrt{ \beta} \left(c - \frac{b r^2}{\sqrt{ \beta}} \right)
\end{equation}

since we have assumed $\left(c - \frac{b r^2}{\sqrt{\beta}} \right) > 0$, $M_5$ must be positive and depends on radial coordinates only and it should be constant in $r$-constant hypersurface.\\
(ii) Again with the help of  equations \eqref{eq:21}, \eqref{eq:22} and \eqref{eq:39} we  calculate

\begin{equation}  \label{eq:45b}
M_5 = \frac{3}{2} dr^2
\end{equation}
 this  depends also on $r$ but it conserves for $r$-constant hypersurface.\\

 (iii) Next from equations \eqref{eq:25}, \eqref{eq:26} and \eqref{eq:40} we derive

 \begin{equation}  \label{eq:45c}
M_5 = 3B \sqrt{2 \beta}
\end{equation}

which is a constant quantity.

But for a $4D$ observer the effective $4D$ matter is given by $\rho e^{3\lambda} = M_4$ in each case,
such that

(i)
\begin{equation}  \label{eq:46a}
M_4 (t) = \frac{3}{2} \sqrt{ \beta} \left(c - \frac{b r^2}{\sqrt{ \beta}} \right) e^{-\mu}
\end{equation}

(ii)
\begin{equation}  \label{eq:46a}
M_4 (t) = \frac{3}{2} dr^2 e^{-\mu}
\end{equation}
and
(iii)
\begin{equation}  \label{eq:46a}
M_4 (t) = 3B \sqrt{2 \beta}  e^{-\mu}
\end{equation}
respectively.

Since for a realistic model the extra dimension shrinks, the above equation tells us
that although the overall $(4 + 1)D $ matter remains conserved for any r-constant hypersurface there will be matter leakage from the internal space onto the effective $4D$ world. So it provides an alternating method of matter creation in $4D$ spacetime without the assumption of an
extraneous field ~\cite{der} as also without breaking any conservation principle in physics

\subsection{ Accelerating universe }

According to the current cosmological observations of type Ia supernovae ~\cite{sm,res},
the universe is expanding at an accelerating rate. It is commonly accepted that the dark
energy candidate $\Lambda$  in Einstein’s field equations is responsible for the cosmic acceleration of the universe. So we have to check whether our model is accelerating or not. The simplest way to check accelerating phenomenon, we now calculate the  deceleration parameter of three space using equation \eqref{eq:22} which is given by

\begin{equation}  \label{eq:47}
q = - \frac{1}{H^2}\left(\ddot{\lambda} + \dot{\lambda^2} \right)= \frac{2 - \cosh^2 2 \beta t}{\cosh^2 2 \beta t}
\end{equation}
 where $H$ is the Hubble parameter.

\begin{figure}[ht]
\begin{center}
  \includegraphics[width=8 cm]{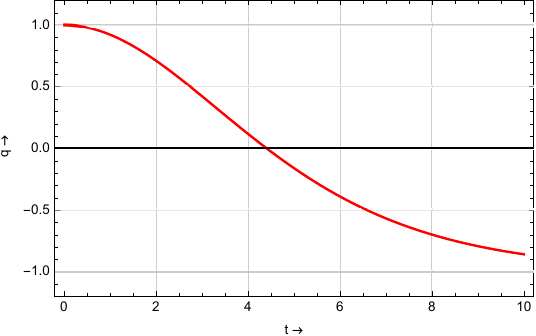}
  \caption{
  \small\emph{The variation of $q$ with $t$ using equation \eqref{eq:47} }\label{fig:qt}
    }
\end{center}
\end{figure}

this gives the flip time $t_f = \frac{0.440687}{\sqrt{\beta}}$, since the value of $\beta $ is very small, represents late acceleration. So it is seen that the usual three space initially expands with deceleration and after flip time ($t_f$), it accelerates which is compatible  with our present observable universe (see fig. \ref{fig:qt}). The flip time depends on the  positive  cosmological constant and it favours for late acceleration. It may be pointed out that this acceleration is not affected by the radial coordinate, since we dealing with the three space only which is a function of time only.

\subsection{ Raychaudhuri equation }

We know that for Einstein's field equation collapse to singularity is inevitable for a realistic matter field, which may be halted with the inclusion of the cosmological constant $\Lambda$ in Raychaudhuri equation~\cite{ray1, ban3,el}. Since our observable universe is currently undergoing a late acceleration and $\Lambda CDM$ model explains the observational facts most closely we have thought it fit to compare our findings with Raychaudhuri equation generalised to $(4+d)$ dimensional spacetime with positive cosmological constant for our irrotational system  as given by

\begin{equation}  \label{eq:48}
\dot{\theta} = -2 \sigma^2 - \frac{1}{3+d}\theta^2 - \frac{8 \pi G}{2 + d}\left\{(1+d)\rho + 3p + dp_5 \right\} + \frac{2}{2+d} \Lambda
\end{equation}

In our case, $d = 1$ and $p = p_5 = 0$, the equation \eqref{eq:48} reduces to

\begin{equation}  \label{eq:49}
\dot{\theta} = -2 \sigma^2 - \frac{1}{4}\theta^2 - \frac{8 \pi G}{3}2\rho  + \frac{2}{3} \Lambda
\end{equation}

where, $\theta = 3\dot{\lambda} + \dot{\mu}$ and $ \sigma^2 = \frac{3}{8}\left(\dot{\mu} - \dot{\lambda} \right)^2$, with $8 \pi G = 1$, the equation \eqref{eq:49} is given by

\begin{equation}  \label{eq:50}
\rho = - \frac{3}{2} \left( 3 \ddot{\lambda} + 3 \dot{\lambda}^2 + \ddot{\mu} + \dot{\mu}^2\right) + \Lambda
\end{equation}

It is seen that the expressions of matter density $\rho$ in the equations \eqref{eq:30}, \eqref{eq:37} and \eqref{eq:40} are identical with  results obtained from the equation \eqref{eq:50} for different models discussed previously, i.e., our results are confirmed by the Raychaudhuri equation.

\subsection{ Deceleration parameter }

The deceleration parameter for $(d+3)$ dimensional spacetime is given by
\begin{equation}  \label{eq:51}
q = \frac{\dot{H} + H^2}{H^2} = -1 - (d + 3) \frac{\dot{\theta}}{\theta^2}
\end{equation}

where $H$ is denoted as Hubble parameter. In our case $d = 1$ and the equation \eqref{eq:51} reduces to

\begin{equation}  \label{eq:52}
q =  -1 - 4 \frac{\dot{\theta}}{\theta^2}
\end{equation}

Now using equations \eqref{eq:49}  and \eqref{eq:52} we get other expressions of deceleration parameter which is given by

\begin{equation}  \label{eq:53}
q = \frac{1}{\theta^2} \left(8 \sigma^2  + \frac{8}{3} \rho - \frac{8}{3} \Lambda \right)
\end{equation}

\begin{figure}[ht]
\centering \subfigure[Model-1:Case-I and $\left(c - \frac{br^2}{\sqrt{ \beta}}\right) > 0$
]{
\includegraphics[width= 6 cm]{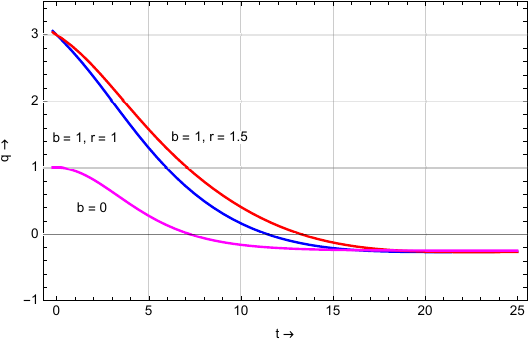}\label{fig:qt1} } ~~~\subfigure[Model-1: Case-II and $\left(c - \frac{br^2}{\sqrt{ \beta}}\right) >0$
 ]{
\includegraphics[width= 6 cm]{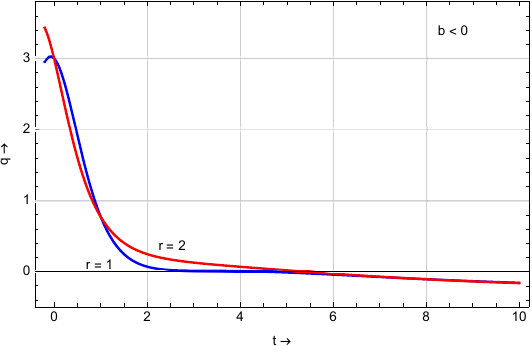}\label{fig:qt2} }~~~\subfigure[Model-2: $ c  >  \frac{b \sqrt{2}}{\beta^{\frac{3}{2}}}$
]{
\includegraphics[width= 6 cm]{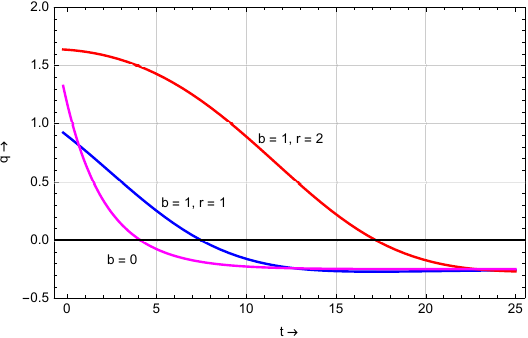}\label{fig:qt3} }\label{fig:figure8}~~~~~~~~~~~\caption[Optional
caption for list of figures]{\emph{ $q$ vs $t$.  }}
\end{figure}

Using the equation \eqref{eq:53} and the solutions of different model discussed above we may calculate the different expressions of the deceleration parameter $q$. As these expressions of $q$ are very difficult to handle analytically, we take recourse to its graphical representations considering the values of constants are parameters which are shown in the fig. \ref{fig:qt1}, \ref{fig:qt2}, \ref{fig:qt3}. It is shown that our universe initially decelerates and after flip it accelerates in each case discussed above. This feature is in agreement with the deceleration parameter of three space in equation \eqref{eq:47} discussed previously. But It is noticed that the flip time depends upon the radial coordinates  also for all the models which is different from the nature of three space  deceleration parameter. This is because we have considered the inhomogeneity in extra space only. Further it follows from the graphs \ref{fig:qt1}, \ref{fig:qt3} that all the deceleration parameters converge at late time for both $b = 1$ and $b = 0$ (homogeneous case). For $b < 0$, all the deceleration parameters coincide at late stage of the evolution of the universe (fig. \ref{fig:qt2}).

\section{Discussion }

We have studied an inhomogeneous dust dominated model with positive cosmological constant $\Lambda$ in higher dimensional spacetime.  We derive the solutions using field equations with the help of the equation of state ($p = 0$). Depending on the value of the integration constant $k$ in equation \eqref{eq:14}, we get two types of solutions denoted by Case I($k = 1$) \& Case II ($ k = 0$). Again in Case I we have derived  two  hyperbolic solutions of $e^{\mu}$ and in Case II  we get exponential solution discussed earlier in the previous work of Banerjee \textit{et al}~\cite{ban2}. This work is a sort of  generalisation of our previous work ~\cite{ban1}.

Here in  all possible cases, standard three space expands indefinitely but the extra space contracts, starting from infinity at $t = 0$ with one minima and then expands indefinitely. We assume that it stabilises by some stabilisation mechanism and so expansion of extra dimension does not occur here. We have calculated the curvature and found that for positive curvature, density $\rho$ reduces asymptotically with time.  The extra dimensional pressure  $p_5 $ automatically vanishes  for all cases. The arbitrary constant $b$ determines the measure of curvature of the 4D space and  depending on signature of $b$, we get positive, negative or flat space according as $b >, < , 0$.  Here it is seen that $R^{*(4)} \rightarrow 0$ as $ t \rightarrow \infty$ implying flat surface. It is found that dimensional reduction occurs due to inhomogeneity. Further it is noticed that the constant $ b$ is the measure of curvature and it accounts for inhomogeneity also. So it may be pointed out that the curvature is a cause of dimensional reduction of extra space for $\Lambda > 0$. It is also seen that when $b = 0$, the inhomogeneity of extra space disappears and it expands indefinitely. In this context, we refer to  the work of Tosa ~\cite{tosa} where it is shown that dimensional reduction is not possible for homogeneous model when $\Lambda > 0$. So $b$ is the constant who changes the total scenario. As the inhomogeneity parameter $b$ tends to zero the coordinate boundary reaches more and more from the point of symmetry and the cosmology mimics increasingly a $5D$ homogeneous model. It is to be noted that the dimensional reduction is affected by the spatial co-ordinate $r$ also. Here 3D scale factor expands indefinitely. For $k=1$, we have discussed the findings for both positive and negative curvature. When we put $k =0$, we get exponential solution and it also reduces to homogeneous  space for $b = 0$.

An encouraging aspect in our model is that the very high value of entropy per baryon observable at present universe may be interpreted as due to dimensional reduction of extra space. It is also observed that as the extra dimension shrinks, the total $5D$ world matter is conserved, and there will be matter leakage from the internal space onto the effective $4D$ world. We have investigated  whether three space accelerate at late time and it is found that the usual three space initially expands with deceleration and after a certain  \textit{flip time } ($t_f$), it accelerates which is compatible  with our present observable universe. The flip time depends on the  positive  cosmological constant and it favours late acceleration. It may be pointed out that this acceleration is not affected by the radial coordinate, since we are dealing with the three space only which is a function of time only. We confirm  our results also obtained for different models with those obtained from the Raychaudhuri equation. The deceleration parameter was studied from Raychaudhuri equation where our universe decelerate initially and after flip it accelerates and flip time depends on  the  radial coordinates also for both Model-1 \& Model-2.

The present work suffers from a serious shortcoming in that the extra dimension initially decreases with time.
 It attains a minimum to expand indefinitely with time once again. To circumvent this difficulty, we assume that the extra dimension stabilises by some quantum mechanical phenomena when it minimises and there is no scope of expansion of extra dimension.  Secondly, we have so far not attempted to  constrain the free model parameters with the help of observational data as it is customary in relevant works in this field. The issue of compatibility of the obtained results with observational data may be addressed in our future work.
\vspace{0.2 cm}

\textbf{Acknowledgement:} We sincerely acknowledge the suggestions of an anonymous Referee which led to a significant improvement of the original version.

\textbf{Data Availability Statement:} No Data associated in the manuscript.

\end{document}